\documentstyle[prl,aps,amsmath]{revtex}
\twocolumn

\begin{document}
%\input epsf
\draft
\title{Towards the elusive Efimov state of the $^4$He$_3$ molecule
  through a new atom-optical state-selection technique}  
\author{Gerhard C.~Hegerfeldt}
\address{Institut f\"ur Theoretische Physik, 
  Universit\"at G\"ottingen, 
  Bunsenstr.~9, 37073 G\"ottingen, Germany}
\author{Thorsten K\"ohler}
\address{Max-Planck-Institut f\"ur Str\"omungsforschung, 
  Bunsenstr.~10,
  37073 G\"ottingen, Germany} 
\maketitle

\begin{abstract}
  Excited states and excitation energies of weakly bound systems,
  e.~g.~atomic few-body systems and clusters, 
  are difficult to study experimentally. For this purpose 
  we propose a new and very 
  general atom-optical method which is based on inelastic 
  diffraction from transmission gratings. The technique is 
  applicable to the recently found helium trimer molecule 
  $^4$He$_3$, allowing for the first time an investigation 
  of the possible existence of an excited trimer state and 
  determination of its excitation energy. This would be of 
  fundamental importance for the famous Efimov 
  effect.
\end{abstract}


Already in 1935 Thomas
\cite{Thomas}
discovered a surprising property of three-body systems.
He considered short-range two-particle potentials which 
supported just one single bound state with an arbitrarily 
weak binding energy. He then found that for the three-body 
system there could exist a much more tightly bound ground
state and that the binding energy could approach $-\infty$
when the range of the two-particle potential approached zero.
Thirty-five years later Efimov
\cite{Efimov}
obtained a striking generalization of this result. 
He predicted
that when one {\em weakened} the two-body interaction 
the number of 
three-body excited states could increase to infinity.
An excited state which appears under weakening of the
two-particle potential is called an Efimov state.
Conversely, under strengthening of the two-particle potential 
an Efimov state disappears into the continuum.
Intuitively one can understand the Efimov effect in a 
three-body system by imagining a weakly bound, and therefore
spatially very extended, two-particle subsystem. This subsystem
can then act on the third particle with a force whose range 
is given by the spatial extent of the subsystem
\cite{Efimov1990}.
This range therefore increases more and more when the 
two-particle potential is decreased.
As opposed to a short-range potential 
\cite{Schwinger},
however, a long-range potential can have infinitely many 
excited states. 

Whether the Efimov effect does occur in nature 
is still an open question.
In nuclear physics no generally accepted examples 
of Efimov states have been found
\cite{Gloeckle}. 
For systems of neutral
atoms, however, they might exist, and an excellent candidate 
is the helium trimer, $^4$He$_3$, since the dimer, $^4$He$_2$, is 
believed to have an extremely weak binding energy ($\approx -1.3$ mK)
and no excited states. 
A few years ago $^4$He$_2$ has been observed
by Luo et al.~\cite{LuoDimer}
and independently by
Sch\"ollkopf and Toennies
\cite{STScience},
who also observed $^4$He$_3$.
There has been a lot of theoretical work on 
the existence of an Efimov state in the helium trimer,
with sometimes conflicting results
\cite{LimEfimov,LimUPA,Gloeckle,Greene,Gianturco,Fedorov}.
The overall theoretical picture presently indicates the existence
of a ground state and a single excited state, denoted here 
by $^4$He$_3$ and $^4$He$_3^*$, respectively. 
The latter is believed to be an Efimov state.
Both are $s$ states with respective 
energy around $E_0=-0.1$ K and $E_1=-2$ mK
\cite{Sandhas}.
Experimentally, $^4$He$_3^*$ has not yet been seen
\cite{STParis}.

The Efimov property as well as the very existence of the excited 
state $^4$He$_3^*$ depend sensitively on the detailed form of the
two-atom interaction, 
and even small retardation corrections to the
potential can significantly affect the  $^4$He$_2$
binding energy and bond length
\cite{LuoRetardation}.
A precise knowledge of this potential is also necessary for 
understanding liquid helium droplets 
\cite{droplets}
and superfluidity.
Therefore conclusive experimental evidence of an excited state
$^4$He$_3^*$ 
and determination of its binding energy would not only
be a crucial step towards establishing the existence of an Efimov 
state but would also give important information on the 
He-He potential. 

Here we propose a new and very general method to both detect
and select an excited state of $^4$He$_3$, or of other systems, 
as well as to determine its excitation energy. 
Excitation can be achieved by scattering a
beam from a solid surface 
\cite{Bertino}
or, in our case, from many small surfaces.
Taking for the latter the bars of a microfabricated transmission 
grating 
\cite{Smith}
one can achieve excitation and separation at the same 
time, as will now be shown.
 
The state-selective property
stems from two
conservation laws. If the incident molecular center of mass momentum
is denoted by $P'$ and the final one
by $P$, energy conservation implies
\cite{phonons}
\begin{equation}
  \label{energybalance}
  P^{\prime 2}/2M=P^2/2M+\Delta E_{\rm int}
\end{equation}
where $M$ is the molecular mass and $\Delta E_{\rm int}$ accounts for
a possible change of the internal molecular state. 
The other conservation law comes from the discrete translational 
invariance of the grating in the 2 direction
(period $d$, cf.~Fig.~\ref{Fig1}). This implies the conservation 
of the initial momentum component $P_2'$ up to a
reciprocal lattice vector
\cite{Ashcroft}, 
i.~e.
\begin{equation}
  \label{momentumconservation}
  P_2=P_2'+n2\pi\hbar/d
\end{equation}
where $n=0,\pm 1,\pm 2,\ldots$.
With $P_2=P\sin\varphi$ (see Fig.~\ref{Fig1}) this
yields a relation between the angle of incidence $\varphi'$ and the 
allowed angle, $\varphi_n$, of the $n$-th order diffraction peak.
With $\lambda'=2\pi\hbar/P'$ and $\lambda=2\pi\hbar/P$ the 
initial and final de Broglie wave length, the relation can be 
written, after a short calculation, as
\begin{equation}
  \label{sinphin}
  \sin\varphi_n=\frac{\lambda}{\lambda'}
  \sin\varphi'+n\frac{\lambda}{d}.
\end{equation}
Eq.~(\ref{sinphin}) holds for any molecule-grating
interaction potential.
A wave theoretical interpretation of Eq.~(\ref{sinphin})
can be given as follows 
(see Fig.~\ref{Fig1}). 
First, a wave with angle of incidence 
$\varphi'$ is {\em refracted} in the plane $A$, with a change of
wave length from $\lambda'$ to $\lambda$. The refraction angle
$\varphi_0$ and the incidence angle $\varphi'$ are related as in
Snell's law
\cite{Jackson}
through $\sin\varphi_0/\sin\varphi'=\lambda/\lambda'$.
Secondly, in the plane $B$, the new wave of wave length
$\lambda$ is {\em diffracted} by the slits as in classical optics
\cite{BornWolf}. 
Combining this with Snell's law gives
Eq.~(\ref{sinphin})
\cite{refraction}.

For an elastic process $\lambda$ coincides with the initial
$\lambda'$ and then Eq.~(\ref{sinphin}) 
reduces to the condition for grating diffraction of a de 
Broglie wave as obtained from classical optics
\cite{BornWolf}. 
But if the molecule is excited by the 
interaction with the grating, $\lambda$ differs from $\lambda'$
by a factor of
$1/\sqrt{1-(E_1-E_0)/E_{\rm kin}'}\cong 1+(E_1-E_0)/2E_{\rm kin}'$
where $E_{\rm kin}'\equiv P^{\prime 2}/2M$ is the initial
kinetic energy.
For $^4$He$_3$ three different kinds of 
processes can occur, namely elastic scattering 
($^4$He$_3\rightarrow ^4$He$_3$),
excitation ($^4$He$_3\rightarrow ^4$He$_3^*$)
and breakups ($^4$He$_3\rightarrow ^4$He$_2+ ^4$He or
$^4$He$_3\rightarrow ^4$He$+ ^4$He$+ ^4$He) where the latter have 
diffuse scattering angles.
For the elastic case ($\lambda=\lambda'$) the diffraction term
$n\lambda/d$ in Eq.~(\ref{sinphin}) has been used 
previously to separate molecules of different mass
\cite{STScience,Pritchard}.

In order to separate the equal mass particles
$^4$He$_3$ and $^4$He$_3^*$
we propose here to use the first term in 
Eq.~(\ref{sinphin}) (Snell's law) and its 
dependence on the incidence
angle as follows. For normal incidence of $^4$He$_3$ the low order 
diffraction-peak angles of $^4$He$_3$ and $^4$He$_3^*$ 
differ by micro radians only and are 
practically indistinguishable ($\varphi'=0$ in 
Eq.~(\ref{sinphin})),
but by rotating the grating ($\varphi'\neq 0$) the peaks will
separate. This allows the identification of $^4$He$_3^*$
and determination of $E_1-E_0$. For example, the zeroth order
$^4$He$_3^*$ diffraction angle gives
\begin{equation}
  \label{energydifference}
  E_1-E_0=\frac{P^{\prime 2}}{2M}
  \left(
    1-\frac{\sin^2\varphi'}{\sin^2\varphi_0}
  \right).
\end{equation} 

To quantitatively check the feasibility of our proposal
we have calculated diffraction patterns of a $^4$He$_3$
beam ($^4$He$_3\rightarrow ^4$He$_3$,
$^4$He$_3\rightarrow ^4$He$_3^*$) 
incident under various angles
on a 100 nm period silicon nitride (SiN$_x$) 
transmission grating and for a typical nozzle temperature of 6 K 
\cite{massselection,Scoles}. 
For this we have applied the approach 
of Refs.~\cite{HeKoeI} and 
\cite{HeKoeII} 
to an incident three-body system with the trimer wave 
functions of Fig.~\ref{Fig2}. 
The wave functions have been obtained by means of the momentum space
Faddeev approach and the unitary pole approximation
(see e.~g.~Ref.~\cite{LimUPA}) 
using the Tang, Toennies, Yiu (TTY) potential
\cite{TTY} and they are sufficiently accurate to yield 
$E_0=-0.1$ K and $E_1=-2.1$ mK, comparable to the results of the
adiabatic hyperspherical approach in Ref.~\cite{Greene}. 
In the calculation of the diffraction pattern in the 
Fraunhofer regime
the trapezoidal cross section 
of the grating bars with a wedge angle of $\beta=9^\circ$
(see Fig.~\ref{Fig1}) and the helium-silicon 
nitride van der Waals interaction potential of Ref.~\cite{Grisenti}
have been included. As can be seen in Fig.~\ref{Fig3} a
$^4$He$_3^*$ signal appears in the form 
of side peaks on the elastic 
diffraction peaks for angles of incidence $\varphi' > 10^\circ$.
The energy difference $E_1-E_0$ is very small compared to 
the initial kinetic energy and results in small angle differences 
between $^4$He$_3$ and $^4$He$_3^*$ diffraction peaks. But such
small angle differences are easily 
resolvable in present-day experiments. 

Fig.~\ref{Fig4} reveals the role played by the attractive 
van der Waals interaction between the molecules and the grating 
material. While the elastic processes qualitatively 
follow the predictions
of classical optics, the total transmitted $^4$He$_3^*$ intensity
($^4$He$_3\rightarrow ^4$He$_3^*$) is only
slowly varying over a wide range of incidence
angles $\varphi'$. Therefore, the excitation of the
molecules depends only weakly on the slit projection 
orthogonal to the incidence direction
and will take place mainly at
the slit boundaries. 
If the angle of incidence approaches the wedge angle of the 
grating bars (see Fig.~\ref{Fig1}) the $^4$He$_3^*$
transmission probability reaches its maximum value. 
In this case of grazing incidence on the surface of one side of the
bars the van der Waals interaction strongly affects the excitation 
process and leads to a gain in the total transmitted
$^4$He$_3^*$ intensity.
Theory shows the excitation process to be the more efficient the
larger the overlap region of the two wave functions.

Fig.~\ref{Fig3} indicates that a realization of the proposed
experiment requires the measurement of intensities over five
orders of magnitude. With conventional electron impact 
ionization mass spectrometer detectors 
this accuracy has been achieved so far solely for 
helium-atom beams
\cite{Grisenti}
because the estimated detection efficiency is only $10^{-6}$.
A helium-atom detector with an efficiency of about $10^{-4}$ 
and an improved $^4$He$_3$ source 
are presently under study 
\cite{STprivate}.
One of them, or both, should make it possible 
to measure, for example, the zeroth order
excitation peaks in Fig.~\ref{Fig3} and thus demonstrate the existence
of an excited helium trimer state.

The principle of the proposed 
method of selecting internal states 
and determining the energy difference of the respective 
internal transitions by means of transmission gratings is
not restricted to the particular $^4$He$_3$ molecule. 
An extension to other systems with low-energy internal 
transitions like other helium clusters, Rydberg atoms or 
even more complicated systems
\cite{Arndt}
can be envisaged. Also, the method could
provide pure beam sources of particular internal states 
and might be a useful spectroscopic tool for the investigation 
of internal atomic and molecular transitions which are not 
easily accessible to laser light.

We would like to thank R.E.~Grisenti, W.~Sch\"ollkopf and 
J.P.~Toennies for stimulating discussions.

%%%%%%%%%%%%%%%%%%%%%%%%%%%%%%%%%%%%%%%%%%%%%%%%%%%%%%%%%%%%%%%%%%%
%Figures
\begin{figure}
  \caption{Wave theoretical interpretation for the 
    selection of excited molecules.\label{Fig1}}
\end{figure}
\begin{figure}
  \caption{(a) Hyperradial probability density of the $^4$He$_3$
    ground (solid line) and excited state (dashed line),
    and (b) radial probability density of the $^4$He$_2$
    bound state (solid line) and TTY 
    potential (dotted line).
    These states show the peculiar property that 
    a substantial fraction of the probability densities 
    is located far outside the potential well.\label{Fig2}}
\end{figure}
\begin{figure}
  \caption{Diffraction intensities for elastic 
    ($^4$He$_3\rightarrow^4$He$_3$) and excitation 
    ($^4$He$_3\rightarrow^4$He$_3^*$)
    processes of a ground state helium trimer beam at a nozzle
    temperature of 6 K, incident on a 100 nm 
    period silicon nitride 
    transmission grating under different angles: 
    (a) $\varphi'=0^\circ$, (b) $\varphi'=10^\circ$,
    (c) $\varphi'=20^\circ$ and (d) $\varphi'=30^\circ$.
    \label{Fig3}}
\end{figure}
\begin{figure}
  \caption{Total transmission 
    probabilities of elastic and excitation 
    processes for a ground state helium trimer beam
    diffracted from a 100 nm period silicon nitride 
    transmission grating at nozzle temperatures of 6 K
    and 30 K as functions of the angle of incidence.
    The dashed line is the transmission curve obtained 
    from geometrical optics.\label{Fig4}}
\end{figure}

\begin{thebibliography}{99}
\bibitem{Thomas}
  L.~H.~Thomas, Phys.~Rev.~{\bf 47}, 903 (1935).
\bibitem{Efimov}
  V.~Efimov, Phys.~Lett.~{\bf 33B}, 563 (1970),
  Sov.~J.~Nucl. Phys.~{\bf 12}, 589 (1971).
\bibitem{Efimov1990}
  V.~Efimov, Comments Nucl.~Part.~Phys.~{\bf 19}, 271 (1990).
\bibitem{Schwinger}
  J.~Schwinger, Proc.~Nat.~Acad.~Sci.~(USA) {\bf 47}, 122 
  (1961).
\bibitem{Gloeckle}
  Th.~Cornelius and W.~Gl\"ockle, J.~Chem.~Phys.~{\bf 85}, 
  3906 (1986).
\bibitem{LuoDimer}
  F.~Luo, G.C.~McBane, G.~Kim, C.F.~Giese, and W.R. Gentry,
  J.~Chem.~Phys.~{\bf 98}, 3564 (1993).
\bibitem{STScience}
  W.~Sch\"ollkopf and J.P.~Toennies, Science {\bf 266}, 
  1345 (1994).
\bibitem{LimEfimov}
  T.K.~Lim, K.~Duffy, and W.C.~Damert, Phys.~Rev.~Lett. {\bf 38},
  341 (1977).
\bibitem{LimUPA}
  S.~Nakaichi-Maeda and T.K.~Lim, Phys.~Rev.~A {\bf 28},
  692 (1982).
\bibitem{Greene}
  B.D.~Esry, C.D.~Lin and C.H.~Greene, Phys.~Rev.~A {\bf 54},
  394 (1996).
\bibitem{Gianturco}
  T.~Gonz\'{a}lez-Lezana, J.~Rubayo-Soneira, S.~Miret-Art\'{e}s,
  F.A.~Gianturco, G.~Delgado-Barrio, and P.~Villarreal,
  Phys.~Rev.~Lett.~{\bf 82}, 1648 (1999).
\bibitem{Fedorov}
  E.~Nielsen, D.V.~Fedorov and A.S.~Jensen, J.~Phys.~B: 
  At.~Mol.~Opt.~Phys.~{\bf 31}, 4085 (1998).
\bibitem{Sandhas}
  A.K.~Motovilov, W.~Sandhas, S.A.~Sofianos, and E.A.~Kolganova,
  physics/9910016v2 (1999).
\bibitem{STParis}
  W.~Sch\"ollkopf and J.P.~Toennies, XVII {\em International
  Symposium on Molecular Beams}, Universit{\'e} Paris XI,
  Orsay, France, Book of Abstracts, p.~143 (1997) suggested 
  using the size of $^4$He$_3^*$ 
  (cf.~also Ref.~\cite{Gianturco}).
  They interpreted deviations in $^4$He$_3$ diffraction
  from Kirchhoff theory as a first hint for the possible
  existence of $^4$He$_3^*$.
\bibitem{LuoRetardation}
  F.~Luo, G.~Kim, G.C.~McBane, C.F.~Giese, and W.R. Gentry,
  J.~Chem.~Phys.~{\bf 98}, 9687 (1993);
  F.~Luo, G.~Kim, C.F.~Giese, and W.R.~Gentry,
  J.~Chem.~Phys.~{\bf 99}, 10084 (1993).
\bibitem{droplets}
  S.~Grebenev, J.P.~Toennies and A.F.~Vilesov, Science {\bf 279},
  2083 (1998).
\bibitem{Bertino}
  Rotationally inelastic diffraction peaks of molecules 
  scattered from a metal surface have been 
  observed, for example, 
  in K.B.~Whaley, C.~Yu, C.S.~Hogg, J.C.~Light and 
  S.J.~Sibener, J.~Chem.~Phys.~{\bf 83}, 4235 (1985).
\bibitem{Smith}
  T.A.~Savas, S.N.~Shah, M.L.~Schattenburg, J.M.~Carter,
  and H.I.~Smith, J.~Vac.~Sci.~Technol.~B {\bf 13},
  2732 (1995). The first observations of 
  atom diffraction from transmission gratings
  and from double slits
  have been reported by
  D.~W.~Keith, M.~L.~Schattenburg, H.~I.~Smith, and D.~E.~Pritchard,
  Phys.~Rev.~Lett.~{\bf 61}, 1580 (1988) and 
  O.~Carnal and J.~Mlynek, Phys.~Rev.~Lett.~{\bf 66}, 2689 (1991),
  respectively.
\bibitem{phonons}
  The $^4$He$_3$ kinetic energy is typically of the order of
  45 K (4 meV) and thus amply sufficient for excitation
  of $^4$He$_3$.
  The excitation of phonons in the grating material can be neglected
  in the considered range of small scattering angles and 
  momentum transfers. The $^4$He$_3^*$ state is practically stable
  with respect to spontaneous emission so that the latter decoherence 
  mechanism can also be neglected in the present work.    
\bibitem{Ashcroft}
  N.W.~Ashcroft and N.D.~Mermin, {\em Solid State Physics}
  (Saunders, Fort Worth, 1976).
  This is also easily seen directly from the periodicity of the
  interaction Hamiltonian, cf.~G.C.~Hegerfeldt and T.~K\"ohler,
  Few-Body Systems, Suppl.~{\bf 10}, 263 (1999).
\bibitem{Jackson}
  J.D.~Jackson, {\em Classical Electrodynamics}
  (Wiley, New York, 1975).
\bibitem{BornWolf}
  M.~Born and E.~Wolf, {\em Principles of Optics}
  (Pergamon, London, 1959).
\bibitem{refraction}
  Eq.~(\ref{sinphin}) also holds for light and two
  media. Interestingly though, a lower light velocity in the second
  medium means refraction towards the normal while for matter a lower
  final velocity means refraction away from the normal. This comes
  from the different relation between velocity and wave length.
\bibitem{Pritchard}
  M.S.~Chapman, C.R.~Ekstrom, T.D.~Hammond, R.A.~Rubenstein,
  J.~Schmiedmayer, S.~Wehinger, and D.E.~Pritchard,
  Phys.~Rev.~Lett.~{\bf 74}, 4783 (1995).
\bibitem{massselection}
  A $^4$He$_3$ beam can be obtained by placing a transmission grating
  in a helium molecular beam and extracting the first order elastic 
  peak of $^4$He$_3$.
\bibitem{Scoles}
  The incident molecular velocity is determined by the nozzle
  temperature through 
  $v'=\sqrt{5k_B T/m}$
  where $m$ is the mass of the helium atom, 
  $T$ the nozzle temperature and
  $k_B$ the Boltzmann constant (see
  D.~R.~Miller, in {\em Atomic and Molecular Beam Methods}, 
  edited by  
  G.~Scoles, (Oxford University Press, New York, 1988)).
\bibitem{HeKoeI}
  G.C.~Hegerfeldt and T.~K\"ohler, Phys.~Rev.~A {\bf 57}, 2021
  (1998).
\bibitem{HeKoeII}
  G.C.~Hegerfeldt and T.~K\"ohler, Phys.~Rev.~A, (in press).
\bibitem{TTY}
  K.T.~Tang, J.P.~Toennies and 
  C.L.~Yiu, Phys.~Rev.~Lett. {\bf 74},
  1546 (1994).
\bibitem{Grisenti}
  R.E.~Grisenti, W.~Sch\"ollkopf, J.P.~Toennies, 
  G.C.~Heger\-feldt, and T.~K\"ohler,
  Phys.~Rev.~Lett.~{\bf 83}, 1755 (1999).
\bibitem{STprivate}
  W.~Sch\"ollkopf and J.P.~Toennies, (private communication).
\bibitem{Arndt}
  M.~Arndt, O.~Nairz, J.~Vos-Andreae, C.~Keller,
  G.~van der Zouv, and A.~Zeilinger,
  Nature {\bf 401}, 680 (1999).
\end{thebibliography}
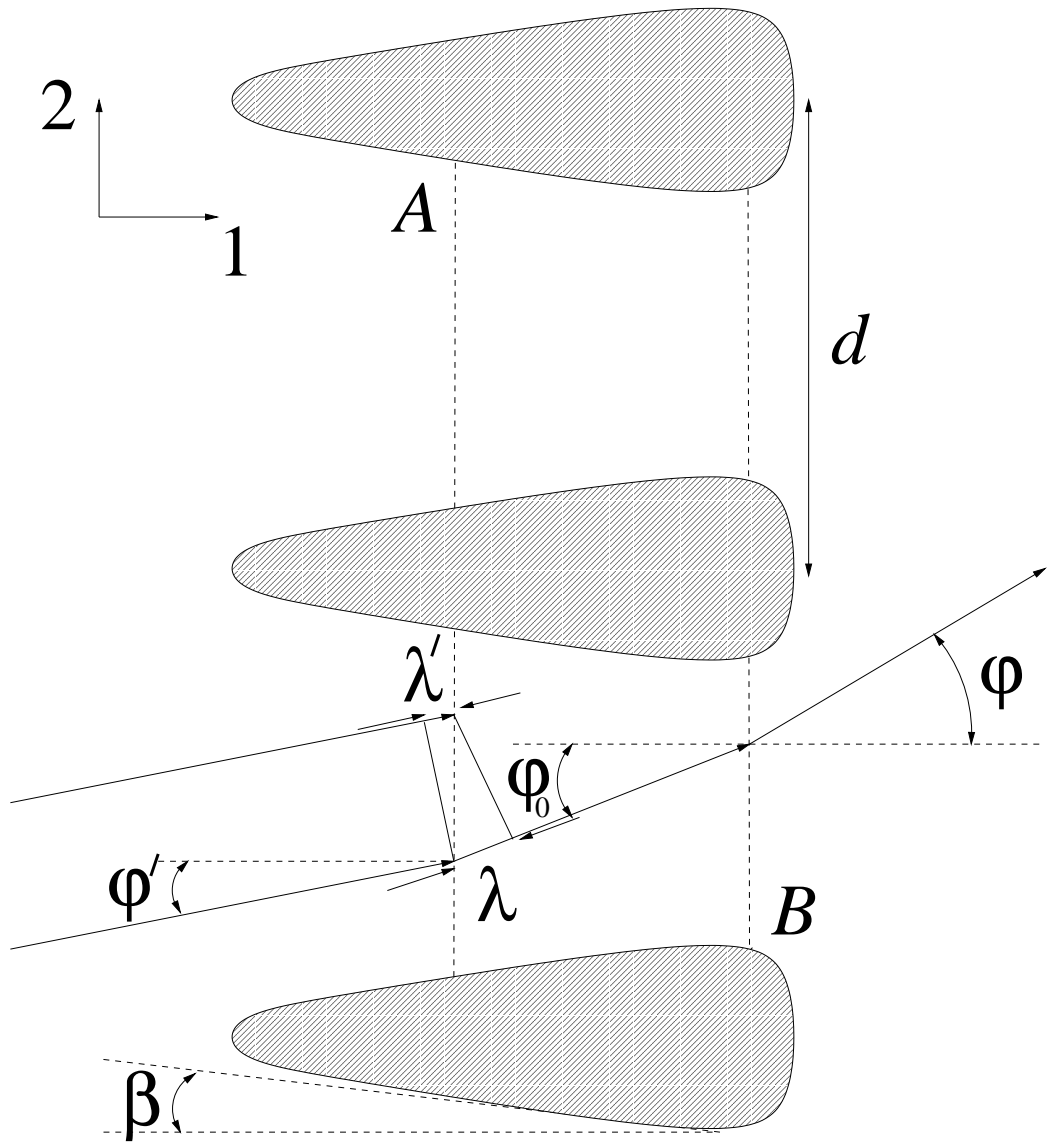
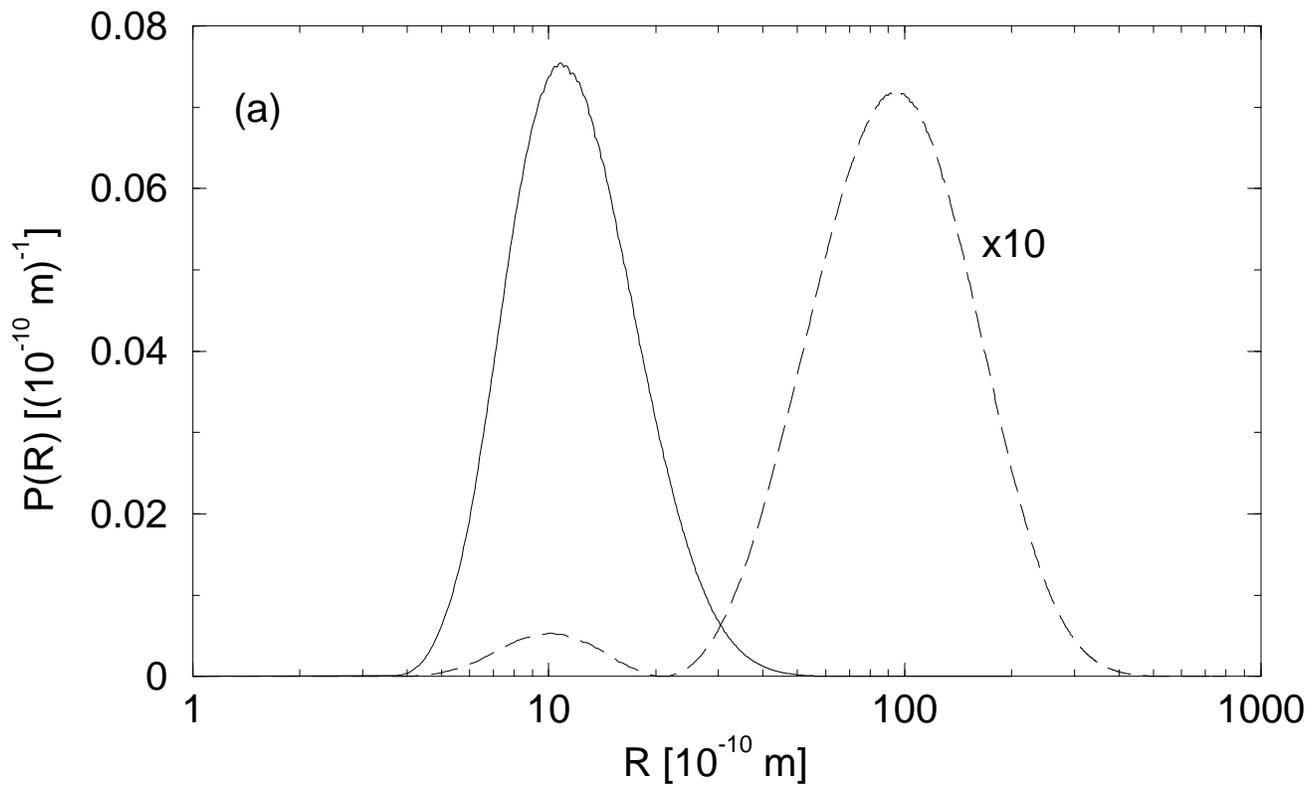
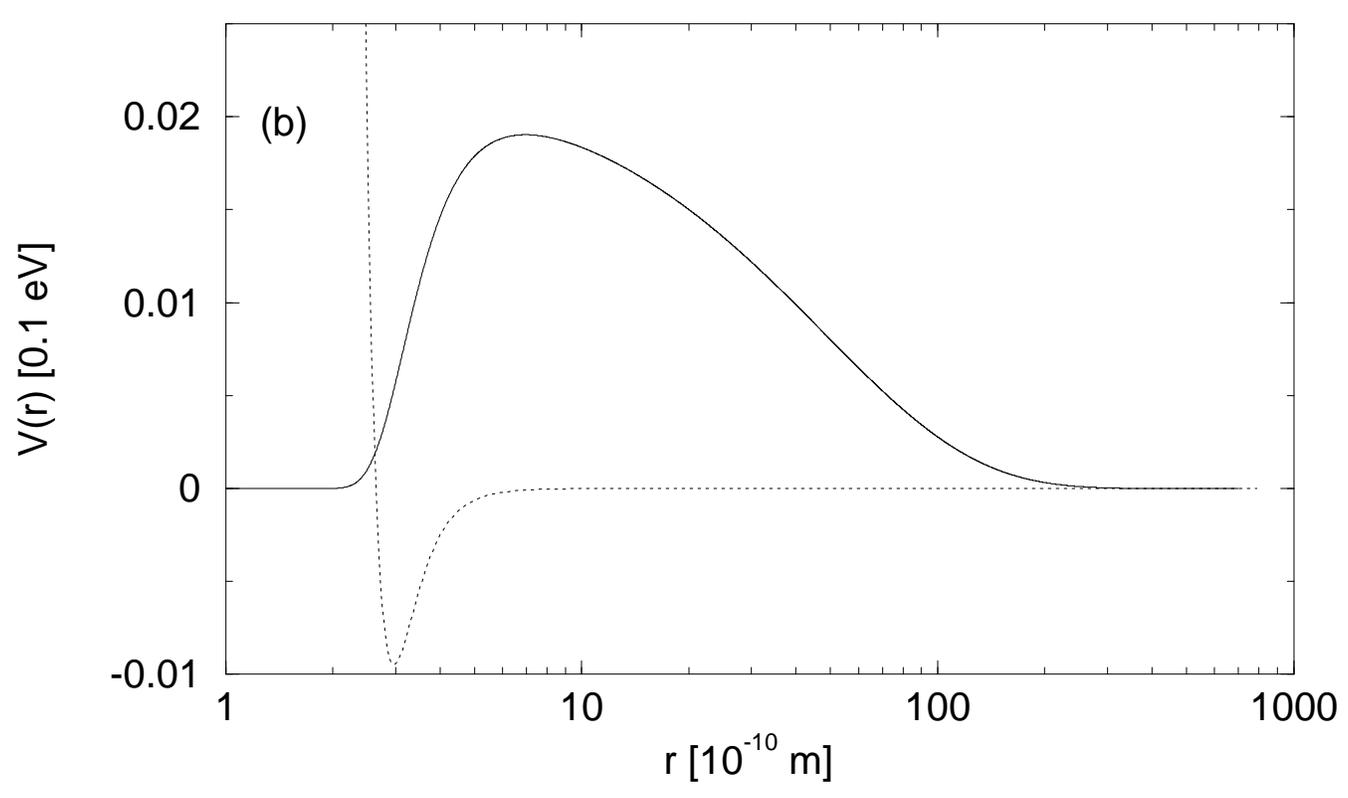
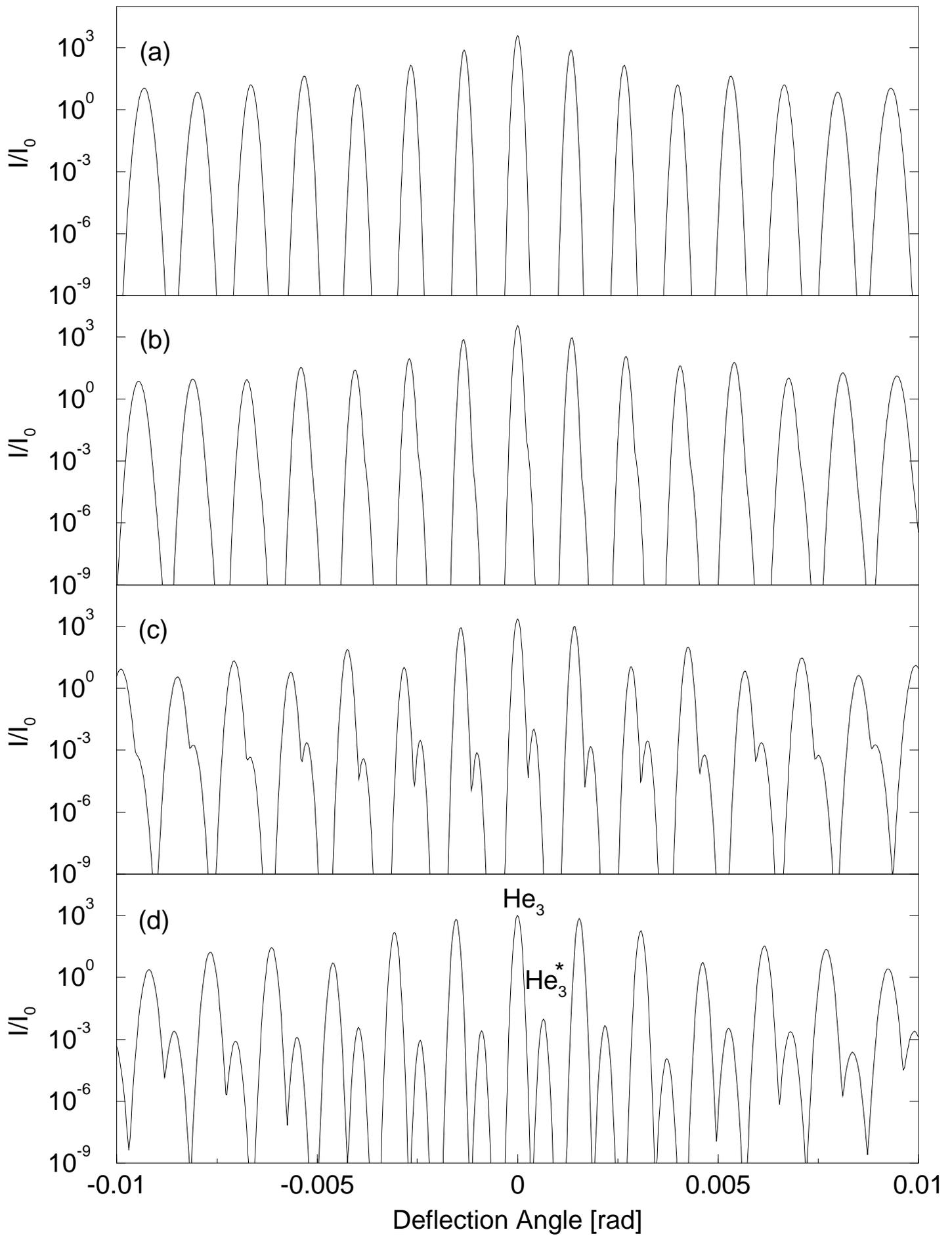
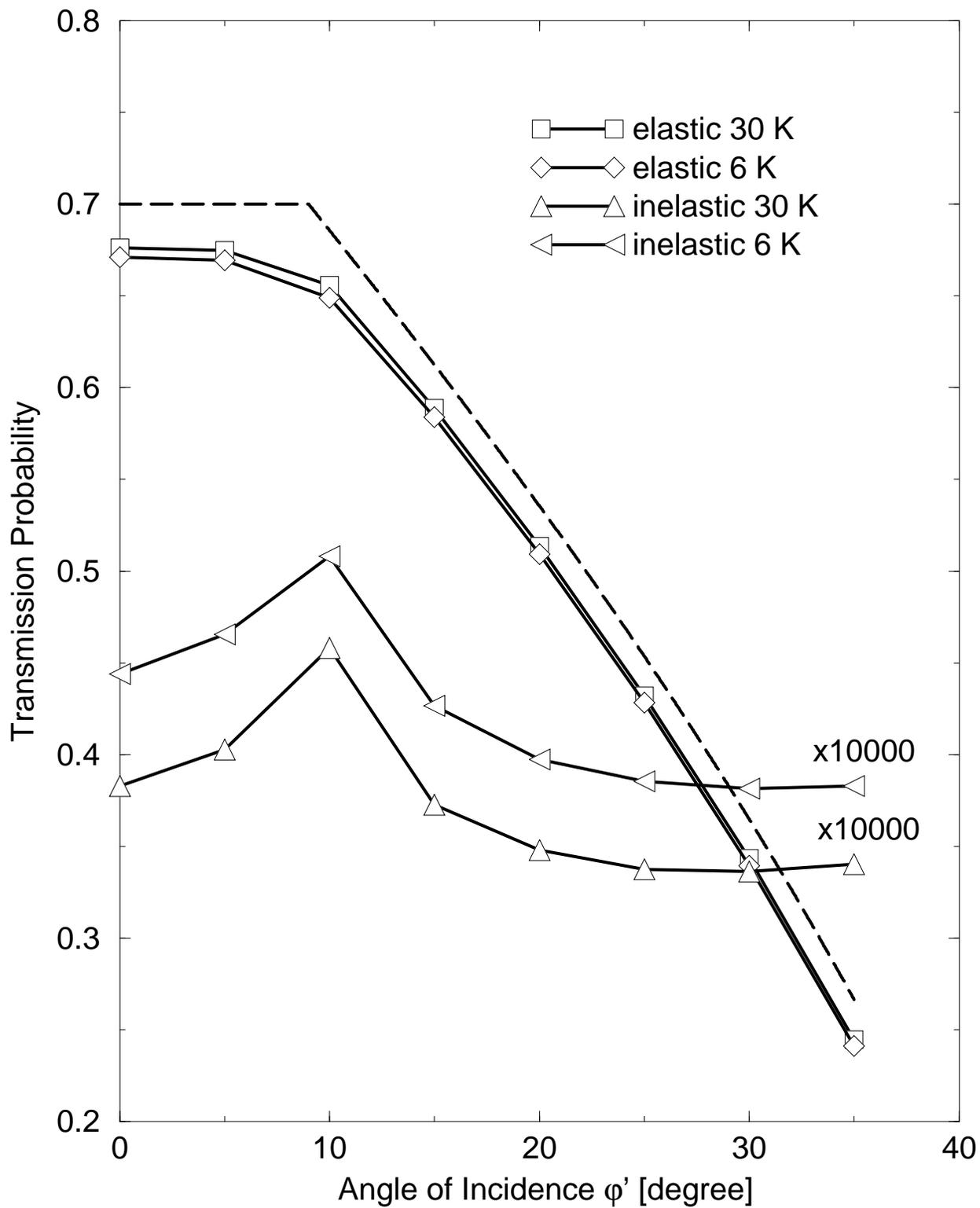
\end{document}